# Emergence of multistability and strongly asymmetric collective modes in two quorum sensing coupled identical ring oscillators.

Edward H. Hellen [a], Evgeny Volkov [b],

AFFILIATIONS
a *Department of Physics and Astronomy, University of North Carolina Greensboro, Greensboro, NC, USA*
b *Department of Theoretical Physics, Lebedev Physical Institute, Leninsky 53, Moscow 119991, Russia*

## Abstract

The simplest ring oscillator is made from three strongly nonlinear elements repressing each other unidirectionally resulting in the emergence of a limit cycle. A popular implementation of this scheme uses repressive genes in bacteria creating the synthetic genetic oscillator known as the Repressilator. Here, we consider the main collective modes produced when two identical Repressilators are mean-field coupled via the quorum sensing (QS) mechanism which is realized via production of diffusive signal molecules. Using the rate of the repressor's production and the value of coupling strength as the bifurcation parameters, we performed analysis of dynamical regimes starting from the two Andronov-Hopf bifurcations of unstable homogeneous steady state, which generate in-phase and anti-phase limit cycles. Pitchfork bifurcation of the unstable in-phase cycle leads to creation of inhomogeneous limit cycles with very different amplitudes in contrast to well-known asymmetrical limit cycles arising from oscillation death. Neimark-Sacker bifurcation of the anti-phase cycle determines the border of an island in two-parameter space containing almost all the interesting regimes including the set of resonant limit cycles, the area with stable inhomogenous cycle, and very large areas with chaotic regimes resulting from torus destruction, period doubling of resonant cycles and inhomogenous cycles. We discuss the structure of chaos skeleton to show the role of inhomogeneous cycles in its formation. Many regions of multistability and transitions between regimes are presented. These results provide new insights into the coupling-dependent mechanisms of multistability and collective regime symmetry breaking in populations of identical multidimensional oscillators.

## <u>Introduction.</u>

Since the work of Huygens and van der Pol it became clear that the design of interactions between nonlinear oscillators can control the types of observed collective dynamical regimes. During the long investigations of the huge number of mathematical models and experimental circuits, it has been shown that both the coupling scheme and the properties of the isolated oscillators are of principle importance for the generation of various modes of collective behaviors. The in-phase, anti-phase, chaotic oscillations as well as their quenching have been observed in populations of appropriately coupled identical oscillators of different nature. During the last two decades the studies of chimeras (coexistence of coherent and incoherent clusters) in homogeneous populations, even of very different natures: phase oscillators [1, 2], chemical oscillators [3], metronome ensemble [4], (see [5] for review) demonstrate strong examples of the dominant role of coupling type in determining the collective behaviors.

However, in many systems the spectrum of possible coupling designs may be limited by the real properties of the isolated oscillators and the environment between them. For example, voltage coupling is the widely accepted standard for electronic circuits while synaptic coupling is prescribed for many neural networks. Diffusion of same-name phase variables is a typical mechanism to couple chemical oscillators embedded in water droplets, but the choice of oils between droplets can control the final collective modes via the selection the rates of diffusion for different variables [6].

Less popular coupling via dissimilar (*conjugate*) variables also may generate different collective behavior. This has been demonstrated, for example, in identical chaotic Rössler oscillators [7], for an electrochemical corrosion model and the Hodgkin–Huxley model for neuronal spiking [8] and for the simple Stuart-Landau model [9]. An extensive review of coupling functions in theory and applications has been recently presented [10].

In many cases the key products of biological oscillators are so strongly located inside the cells that communication in cell populations is realized indirectly via the synthesis of special signal molecules moving between cells. Intercellular communications in bacterial populations are realized via the well-known quorum sensing (QS) system [11] in which the small molecules of *acyl-homoserine lactone* (typically called "autoinducer", AI) diffuse quickly between cells. The effectiveness of such coupling depends on the population density that controls the degree of the autoinducer dilution in the environmental medium. The role of the population density for the synchronization in an ensemble of indirectly coupled biological, chemical and chaotic oscillators has been presented in many publications, e.g. [12-16].

Very recently [17] it has been shown that the indirectly and conjugately mean-field coupled Stuart–Landau oscillators demonstrate the emergence of a spontaneous symmetry breaking oscillatory state, which coexists with a nontrivial amplitude death state. Indirect coupling is realized via the addition of low-pass filter with the parameter named "the mean-field density", which can be considered as a version of the quorum sensing mechanism. Inspired by nature's use of QS coupling in bacterial communication, it became a popular method for managing dynamics in populations of synthetic genetic networks [18-20]. However, the possibilities of QS coupling in the generation of collective regimes are not limited to considering only the population density. There are additional model parameters to consider. It is necessary to point out that the promoter controlling the expression of gene which provides the production of autoinducer and the promoter of target gene which accept the influence of autoinducer, may be different. It is important that due to the current methods of synthetic genetic engineering, the presence of several genes with different promoters in the network and the additional genes for QS coupling support significantly higher flexibility in construction of coupling schemes than that for chemical, biochemical and neuronal oscillators.

After the creation of the synthetic genetic ring oscillator known as the Repressilator [21], it was numerically shown that a population of Repressilators with different periods may be coherently synchronized if the genes coding the elements of QS coupling are implemented to their plasmids [22]. The minimal version of a Repressilator consists of three genes (**a,b,c**) (nonlinear elements) whose protein products (**A, B, C**) repress the transcriptions of each other unidirectionally in a cyclic way (..**A–|B–|C–|A..).** In [22] the autoinducer activates the expression of gene **c** while the production of autoinducer is controlled by the same promoter as that for gene **a**. In order to get other collective regimes, it is enough to change the promoter controlling the autoinducer production.

For example, in [23] gene **b** and the gene controlling autoinducer production share the same promoter. As a result, this coupling changed the basic collective regime for coupled identical Repressilators from in-phase to anti-phase mode and provided the appearance of stable homogeneous and inhomogeneous steady states, chaos, and cluster formation [24, 25]. Recently, the Repressilator has been improved [26, 27], "making it an exceptional precise biological clock" [28]. The electronic analog of the Repressilator was presented in [29, 30].

It is valuable to extend the investigations of coupled Repressilator models for comparison with possible biological experiments and to reveal the mechanisms which control the variability of dynamics in homogeneous populations of ring oscillators. The cooperativity of transcription repression, which is the core process of the Repressilator, is typically described by the Hill function $\sim\alpha/(1 + \mathbf{x}^{\mathbf{n}})$ where **x** is repressor **(A,B,C)** abundance and **n** is the degree of repression cooperativity. The dynamics of the Repressilator are sensitive to the value of **n** which controls the steepness of repression and to **α (A,B,C),** which determines the rates of A,B,C production and therefore, also the amplitudes of the isolated Repressilator variables (see section "Model")**.** It is known that for identical **α (A,B,C)** and identical degradation rates of A,B,C the main condition for the emergence of stable limit cycles is n>2 as was shown in [21]. Rigorous analytical investigations of the Repressilator limit cycle may be found in [31, 32]. The Repressilator's period is controlled by the degradation rate constants of the repressors and it is intuitively expected that stability of the Repressilator's limit cycle is enhanced when the rate constants are the same for all repressors. The effects of non-uniform rate constants on limit cycle stability have been investigated in detail [33].

Recently we [34, 35] investigated the dynamics of two identical 3-dimensional Repressilators QS-coupled as presented in [23] and found many areas of the parameters **α, n, Q** where very rich sets of attractors emerge. It has been shown that without any additional external stimulation, anti-phase limit cycle converts to a two-frequency torus which gives rise to a big family of resonant limit cycles which, in turn, create extended areas with chaos. Despite using identical Repressilators, many periodic windows with inhomogeneous limit cycles have been found inside chaotic areas as well as the unexpected limit cycle with 1:2 winding number (LC1:2) which is stable over a large parameter plane and coexists with other attractors including chaos.

The nature of the LC1:2 was not revealed and our main target here is the detection of the regime whose bifurcation provides the birth of this highly asymmetric cycle and to perform its detailed bifurcation analysis. The structure of this cycle in phase space is different from the well-known inhomogeneous cycles which emerge from coupling dependent inhomogeneous steady states (oscillation death) causing oscillator rotations around different steady states, in stark contrast to the LC1:2. The stabilization of this new type of asymmetric cycle is also far from standard, the period doubling and Neimark-Sacker bifurcations form the borders of stability and transitions to chaos. The phase diagram of regimes associated with LC1:2 shows very large areas with stable inhomogeneous cycles and chaos. Neimark-Sacker bifurcations of the LC1:2 also lead to the production of a family of resonant n:2n limit cycles.

The rest of this work is organized as follows. Section Model contains the description of the model and methods. The bifurcation route to stable strongly inhomogeneous limit cycle and its location on two-parameter phase diagram will be presented in Section Results. The final section is the Discussion.

## **Model.**

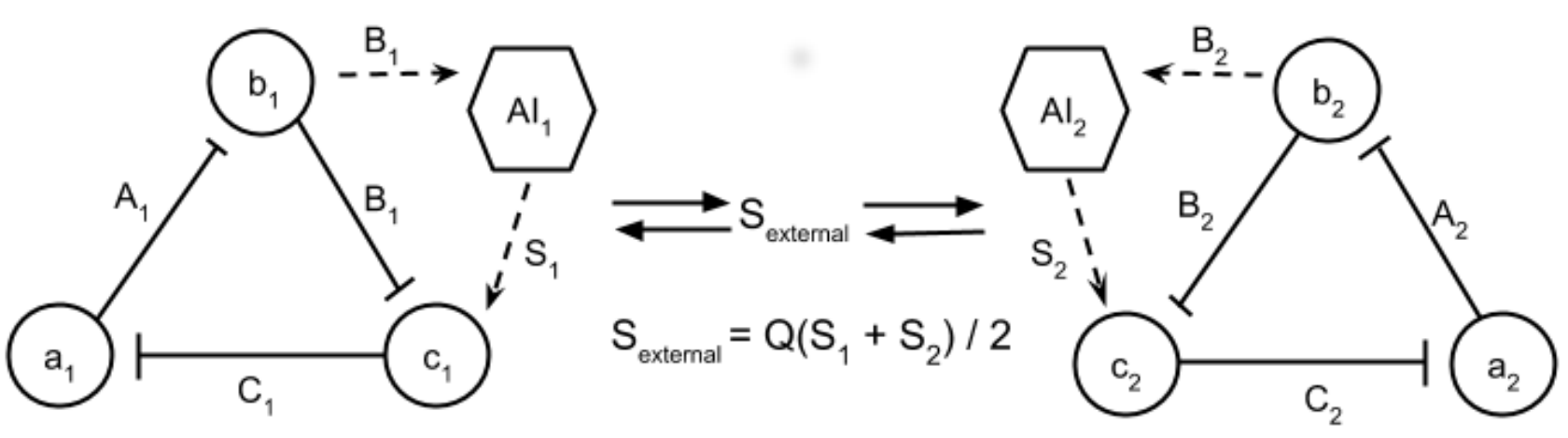

**Fig.1** *The genetic network showing two repressilators coupled via quorum sensing. The autoinducer diffuses through the membrane creating concentrations $S_1$ and $S_2$ inside the cells and $S_{external}$ in the external medium. Lower case (a,b,c) are mRNAs and upper case (A,B,C) are their expressed protein repressors.*

We investigate the dynamics of two identical Repressilators interacting via QS coupling as used previously [23, 24]. Figure 1 shows two repressilators located in different cells and coupled by QS mechanism via the external medium. The three genes in each ring produce mRNAs (a, b, c) which are translated to proteins (transcription factors) (A, B, C), and they impose Hill function inhibition on each other in cyclic order by the preceding gene. The QS feedback is maintained by the autoinducer (AI) molecules which are synthetized inside the cell by a special enzyme LuxI with rate proportional to concentration B with coefficient $k_{S1}$. AI from both cells diffuse through the membrane (coefficient η) and quickly mix in the external environment forming average concentration which influences the activation (rate **κ** in combination with Michaelis function) of mRNA production for protein C, which, in turn, reduces the concentration of protein A resulting in activation of protein B production. In this way protein B plays a dual role of direct inhibition of protein C synthesis and AI-dependent activation of protein C synthesis, resulting in complex dynamics of the repressilator, even for just a single Repressilator equipped by the AI-feedback loop [36]. The original models of a single repressilator [21, 23] used re-scaled dimensionless quantities for rate constants and concentrations of seven variables. We reduce the model for the case of fast mRNA kinetics ((a, b, c) are assumed in steady state with their respective inhibitors (C, A, B), so that $da/dt = db/dt = dc/dt \approx 0$). The resulting equations for the protein concentrations and AI concentration Si are,

$$\frac{dA_i}{dt} = \beta_1\left(-A_i + \frac{\alpha}{1 + C_i^n}\right) \qquad (1a)$$

$$\frac{dB_i}{dt} = \beta_2\left(-B_i + \frac{\alpha}{1 + A_i^n}\right) \qquad (1b)$$

$$\frac{dC_i}{dt} = \beta_3\left(-C_i + \frac{\alpha}{1 + B_i^n} + \frac{\kappa S_i}{1 + S_i}\right) \qquad (1c)$$

$$\frac{dS_i}{dt} = -k_{S0}S_i + k_{S1}B_i - \eta\left(S_i - Q\frac{S_1 + S_2}{2}\right) \qquad (1d)$$

where $i$ = 1,2 for the two Repressilators, parameter **α** accounts for the maximum transcription rate in the absence of an inhibitor, and **n** is the Hill cooperativity coefficient for inhibition. Here we consider the same α and n for all genes. For the quorum sensing pathway $k_{S0}$ is the ratio of the AI decay rate to mRNA decay rate, and as previously mentioned, $k_{S1}$ is the rate of production of AI and **κ** gives the strength of AI activation of protein C production. The diffusion coefficient η depends on the permeability of the membrane to the AI molecule. The second term inside the brackets in Eq. (1d) is the concentration of AI in the external medium, assumed to be a quasi-steady-state combination of the AI produced by both repressilators ($S_1$ and $S_2$), with a dilution factor Q which is controlled by population density that is the origin of the name "Quorum Sensing". $\beta_j$ (j = 1,2,3) are the time scales for repressors turnover which are the ratios of their decay rates to mRNA decay rate. In order to extent work in our previous paper [35] we fix the parameter values n =3.0, κ = 15, $k_{S0}$ = 1, $k_{S1}$= 0.01, η= 2 and start the study of the role of time scales with $\beta_j$=0.5, 0.1, 0.1 taking **α** and Q as basic bifurcation parameters for phase diagrams.

In light of the recent review [10] our synthetic organization of intercellular communications may be classified as “environmental conjugate nonlinear coupling”. Although the change of population density is the general control factor for any realization of QS-coupling, the source of the coupling function flexibility is the freedom in combinations of gene and promoter locations which is inherent to synthetic genetic networks. It does not mean that similar coupling is specifically exclusive to genetics, it is realized in electronic circuits [30, 34, 36] and can serve as a model coupling system for the study of variability in multidimensional biochemical and ecological networks.

Numerical bifurcation continuations are performed with XPPAUT [37] and AUTO-07p [38], time series are calculated by direct integration with 4th-order Runge- Kutta solver, the maps of return times are collected by the section of time series at appropriate value of chosen variable and the spectra of Lyapunov Exponents (LEs) are calculated using method described in [39].

**RESULTS.**

To show the basic bifurcations which are used as starting points in XPPAUT, AUTO [37, 38] numerical calculations of 2-parameter Q-$\alpha$ maps showing dynamical regimes, we present the 1-parameter bifurcation analysis in Fig. 2(a,b). These continuation diagrams for $\alpha$=1200 and 2500 show all the bifurcations of the steady states and the oscillating regimes which emerge from those steady states via Hopf bifurcation. Three types of steady-states and three types of oscillations are found as described below. The low-amplitude homogeneous steady state (HSS) of repressors $B_{1,2}$ is unstable up to high-Q where there are two Andronov-Hopf bifurcations: the first (HB2) corresponds to the emergence of unstable in-phase (IP) limit cycle, and the second (HB1) is where HSS becomes stable and stable anti-phase (AP) limit cycle arises. Further continuation of the HSS consequently encounters the limit point (LP4) bifurcation which destabilizes the HSS and turns its direction, then the splitting of unstable HSS via pitchfork (BP1 in Fig.2(a)) bifurcation creating two inhomogeneous (IHSS) branches: high $B_1$/low$B_2$ or high-$B_2$/low$B_1$. As Q decreases, the amplitude of $B_{1,2}$ in the HSS branch grows reaching the restoration of HSS stability via LP1 bifurcation, while the IHSS branches are stabilized at LP2 bifurcation. High-amplitude HSS is stable up to the large Q values but the stability of IHSS as a function of coupling is limited by LP3 bifurcation which closes the IHSS loop by returning continuation into BP2 bifurcation. Although both HSS and IHSS are stable over noticeable areas of parameters, their locations in phase space are very different from other attractors due to the requirement of high amplitudes of variables $B_i$. Therefore, the presence of HSS/IHSS interferes only weakly with the development of complex oscillating regimes.

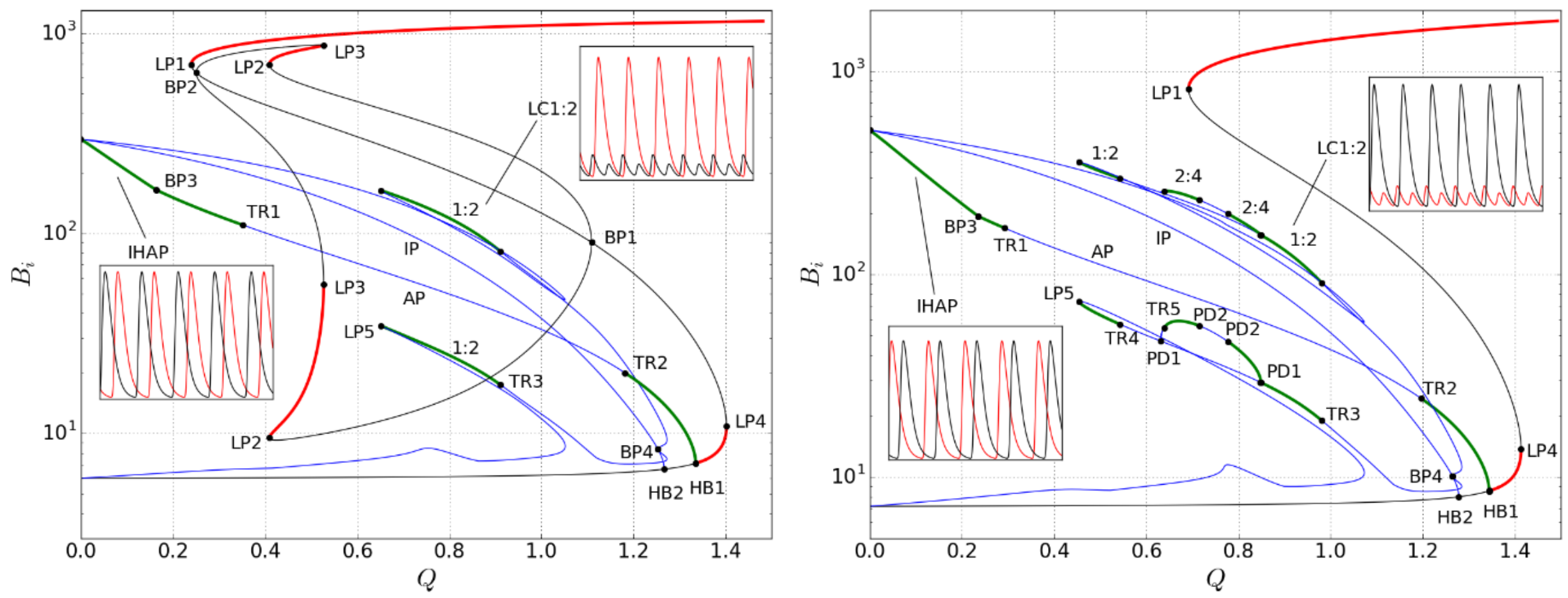

**Fig. 2**. *Amplitudes of variables Bi in system (1) for bifurcation continuations (XPPAUT) of coupling strength. Stable (unstable) steady states are thick red (thin black) and stable (unstable) limit cycles (LC) are thick green (thin blue). TR, LP, BP, HB, are designations for Neimark-Sacker torus, limit point, branch point (pitchfork), Andronov-Hopf bifurcations, respectively. Homogeneous LC branches are the anti-phase (AP) emerging from HB1 and the in-phase (IP) emerging from HB2. Inhomogeneous LC is the two branches with rotating number 1:2. Insets show time-series of the IHAP and the LC1:2. Parameters are* $\boldsymbol{\alpha}$ *= 1200 (a) or 2500 (b),* $\beta_1 = 0.5$, $\beta_2 = \beta_3 = 0.1$.

There are 3 types of broken symmetry displayed in the Fig 2 bifurcation continuations, one from each of the 3 symmetric branches HSS, IPLC, and APLC. The target of interest here are the highly asymmetric LCs which descend from the IPLC via BP4. The other two asymmetric branches are only briefly described below to complete the understanding of the continuation diagrams.

The long in-phase limit cycle branch (IPLC) seen in both continuations emerging from HB2 of the unstable HSS is unstable for its entire Q-interval. However, its pitchfork bifurcation BP4 leads to the strong splitting of oscillation amplitudes of the two oscillators, and with further continuation to its stabilization with unequal rotating number 1:2 via Neimark-Sacker (TR3) bifurcation. This cycle is the LC1:2 whose time series are seen in the insets of Fig. 2(a,b). Further evolution of this highly asymmetric LC for the two values of α are different because for α=1200 the continuation of LC1:2 remains stable down to the final saddle-node bifurcation (LP5), whereas for α=2500, after stabilization by the Neimark-Sacker bifurcation (TR3), the LC1:2 then encounters the cascade of period doubling bifurcations creating chaos in a broad area of the map of regimes, followed by two more Neimark-Sacker bifurcations (TR4,5) before finally getting to the LP5. The TR4,5 are a second source of chaos, in this case via torus destruction. The role of LC1:2 in chaos creation is described below.

The long APLC branch in both continuations in Fig.2(a,b) has two Neimark-Sacker bifurcations (TR1,2) which create a wide unstable region in the central range of the diagram. Most of the interesting dynamics are realized between these bifurcations which will be used here as indication of the region of interest. Symmetry breaking of the APLC at BP3 produces asymmetry of a very different nature as indicated by the Fig. 2 insets showing oscillations of nearly the same amplitude but with phase difference shifted away from the 180° of the symmetric AP.

Figure 2a shows the IHSS emerging from the symmetry breaking of unstable HSS at BP1,2. Under appropriate parameter values Hopf bifurcations appear instead of LP2s creating stable IHLCs accounting for the third type of broken symmetry for these identical oscillators. Although it is absent for α=2500 and the range of parameters in this work, IHSS is the source of inhomogeneous limit cycles as shown for two Repressilators [34] and for small populations of identical Repressilators with other set of parameters [25].

Fig. 3 shows the map of regimes in the Q-α parameter plane for the model Eqs. (1) with $\beta_j$ = 0.5, 0.1, 0.1. The stable LC1:2 occupies large regions of the parameter plane with 2-parameter continuations of its LP, TR3, PD1, PD2, TR3, TR4 and TR5 from Fig.2(a,b). The map shows only selected collective regimes, namely: continuations of special points for the LC1:2, the continuation of TR1/TR2 bifurcations of APLC, and the boundaries of resonant limit cycle with rotation number 5, LC5:5. This cycle has a complex bifurcation structure which will be investigated in future work and here is used only because it coexists with LC1:2 and because these two cycles are partners in the creation of chaos. The more detailed analysis of resonant cycles in this model is presented in [35].

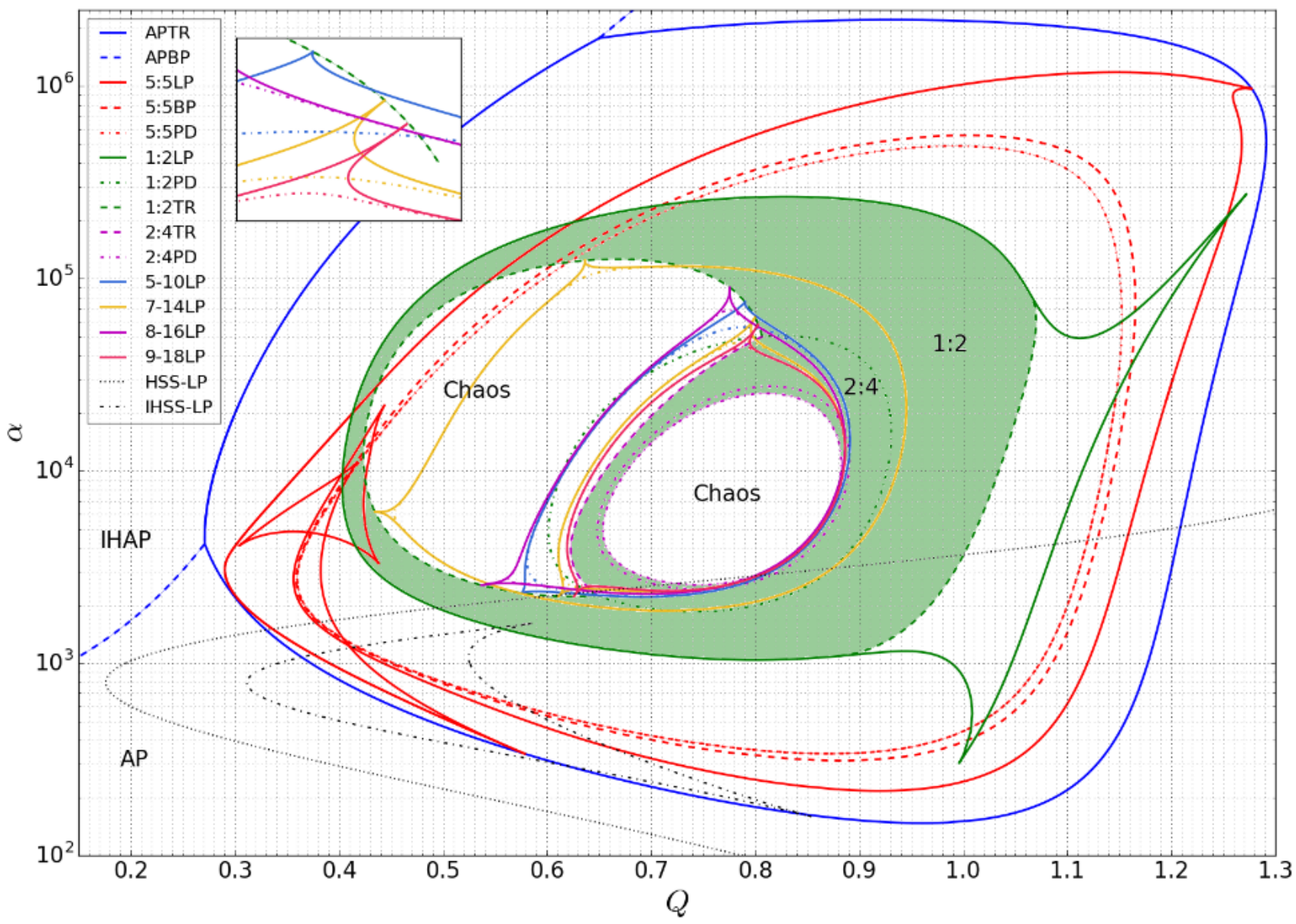

**FIG. 3** *The phase diagram of basic regimes for the system (1) in Q-α plane. For all considered values of **α** the main attractor in the system (1) for low coupling is anti-phase limit cycle which is homogeneous below the line of pitchfork (BP) bifurcation (dashed blue line) but loses symmetry above it (IHAP). Both versions of anti-phase cycle demonstrate Neimark-Sacker bifurcation presented as the closed APTR-line (heavy blue). The red lines are the following bifurcations: saddle-nodes (5:5LP), the pitchfork (5:5BP and period doubling (5:5PD) of resonant cycle LC5:5. Heavy solid (dashed) green lines are the limit point (Neimark-Sacker) bifurcations of inhomogeneous LC1:2. The LP-lines for high-period inhomogeneous cycles LCn:2n have yellow, magenta, rose and blue colors. Insets: the zoomed area with the points of LCn:2n births on the torus and the boundaries of their stability. Large areas of the diagram are occupied by chaos originated from period doubling cascade or via torus destruction. LC1:2 contributes to the two regions labeled "Chaos". Apart from oscillating collective regimes the model (1) may exist in homogeneous (HSS) and inhomogeneous (IHSS) stationary steady states. The boundaries of their stability are pictured by black dotted and dash-dotted lines.*

The map in Fig. 3 suggests three sources of chaos: period doubling cascades for both the LC1:2 and the LC5:5, and the destruction of the inhomogeneous torus. For clarity, only the 1:2's torus destruction and doubling cascade chaos regions are shown in Fig. 3. The PD-line of the LC5:5 indicates its nearby period doubling to chaos. The PD-line forms a closed loop containing much of the LC1:2 and all of its period doubling region. This means that there is a potential for chaos originated from LC5:5 doubling to exist over much of the region occupied by LC1:2 and by chaos originating from the LC1:2 via period doubling and torus destruction. These coexistences and interactions are explored below. In what follows, we explore the map by using Q-continuations, sequential-period (return times) maps, time-series, and Lyapunov exponents (LEs) calculations to examine in detail some of the transitions between regimes and hysteretic regions in the map encountered as Q changes.

As a typical example of analysis, the calculated LE graph in Fig. 4 for α = 5000 shows that the LC1:2 (and its period doublings) and chaos are dominant over a broad range of coupling strength. A noticeable span of inhomogeneous torus is seen following the torus bifurcation TR4. Torus destruction produces growing chaos over a short Q-span followed by a steep jump to a stronger chaos, which we show below is due to the chaos originated from the LC5:5's doubling cascade. Stable LC2:4 exists between its torus bifurcation (TR5) and the left-side of the period doubling cascade to the central island of chaos seen in **Fig. 3.** The first period doubling PD1 on the left side of the island occurs inside the unstable region. The PD2 bifurcations in **Fig. 4** are good indicators of the period-doubling cascades of LC1:2 since the higher LC$n$:2$n$ have resolution too small to be conveniently indicated on the figures. A long span of stable LC occurs after the right-side reverse doubling. At Q = 1.034 stable LC1:2 ends at its torus bifurcation (TR3) resulting in chaos.

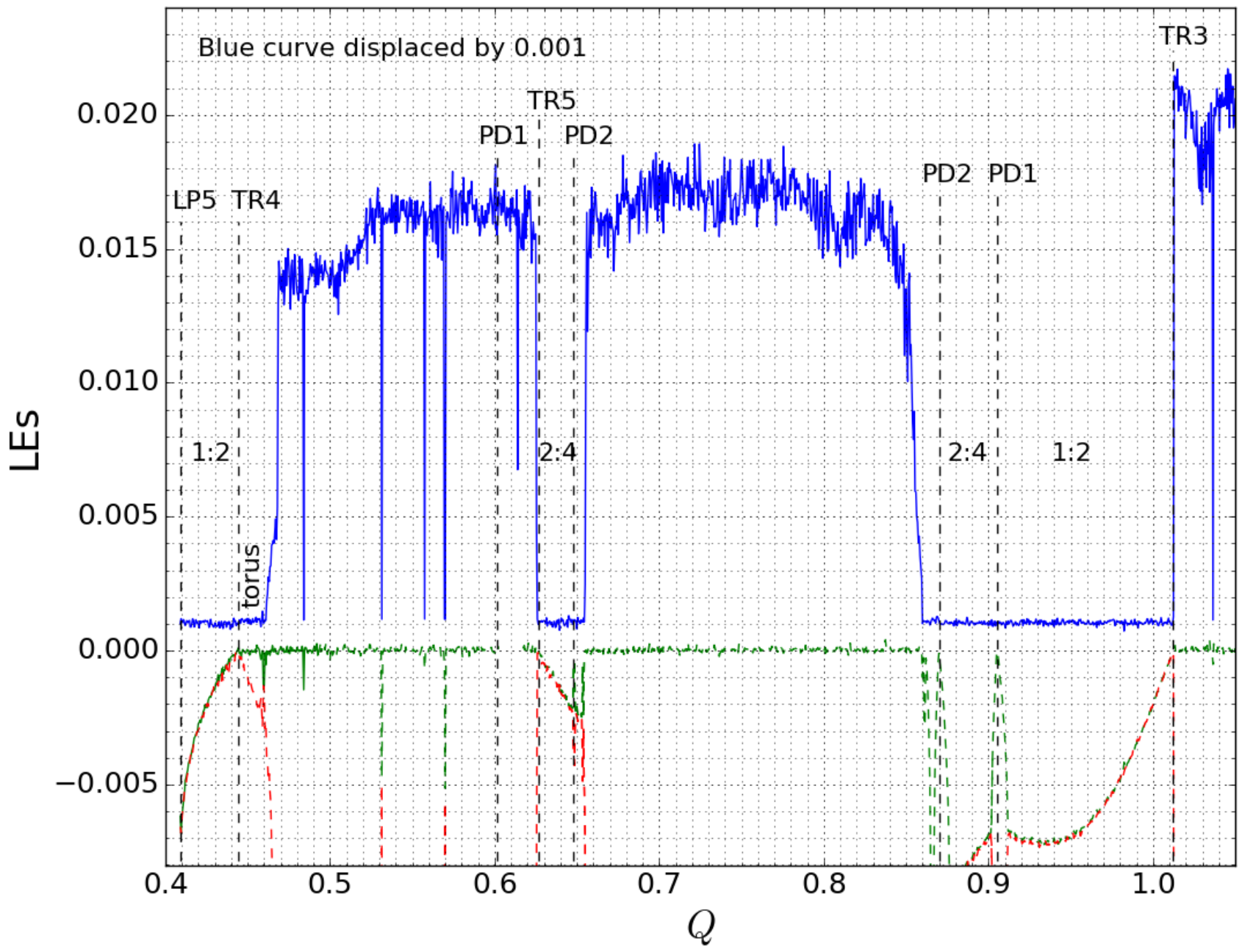

**Fig. 4** *Values of LEs vs Q, starting from LC1:2, α=5000. See description in text.*

The island of chaos created by the period doubling cascade of LC1:2 is inside the larger region defined by the LC5:5 doubling cascade. Below we show that the basin of attraction of the LC1:2 and 2:4 is large enough to capture the chaos originated from the LC5:5 doubling over much of the region. The question then arises about the relative importance of the 1:2 and 5:5 LCs to the structure of the chaos inside the 1:2 doubling cascade. Figure 5a shows a time series taken inside the central island of chaos at Q=0.75 and α=7000. The dominant behavior is highly asymmetric, with variables $B_i$ switching roles as the high/low oscillators. The zoom inset shows that the contribution of LC1:2 to this chaotic time series is quite apparent. For comparison, Fig 5b shows a chaotic time series from the left chaotic region in Fig 3 created by inhomogeneous torus destruction and LC5:5 doubling. As in the central chaotic region, there is switching of IH regions and the influence of 1:2 is seen in the chaos skeleton, however the contributions of other unstable cycles including IP are also apparent. An extensive analysis of the structure of the chaos is beyond the scope of the current work.

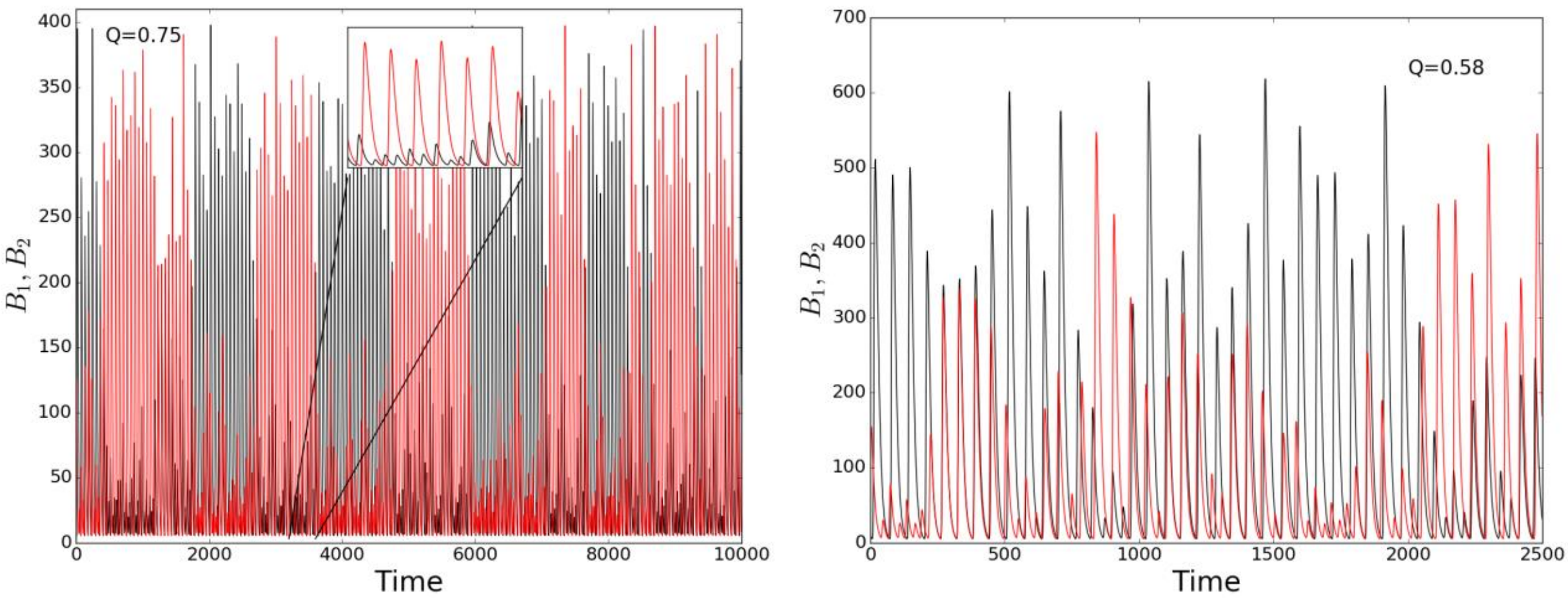

Fig 5. Time series of chaos for α=7000. (a) Q=0.75 (b) Q=0.58. The influence of LC1:2 is clearly present.

Next, we present details of interesting coexistence of regimes and transitions between regimes.

The LE graph for α=5000 in Fig 6 shows a zoom of a region of coexistence indicated by the overlap of LC1:2 and LC5:5 in the map Fig. 3. There is a long interval at low coupling strength with stable LC5:5 fixed by the lines of saddle-node bifurcations seen in Fig 3. Between the 1:2LP and the 5:5LP in Fig 6 the LC5:5 coexists with the inhomogeneous LC1:2. Beyond this 5:5LP the LC5:5 jumps to chaos as indicated by the LE's jump from zero to positive values.

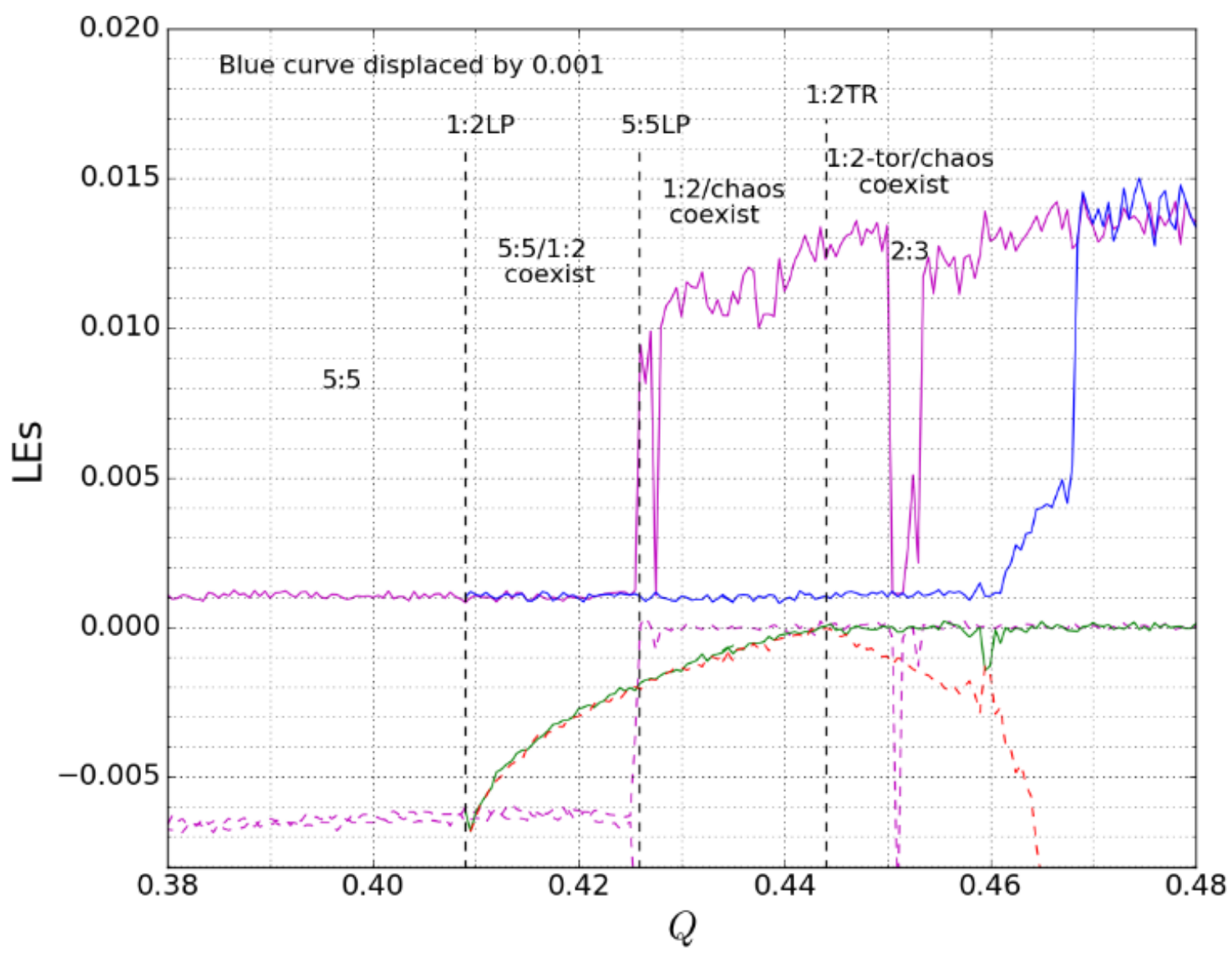

**Fig.** 6 *Evolutions of two families of Lyapunov exponents for trajectories starting from: LC1:2 (blue line) and LC5:5 (magenta line), α=5000.*

Interestingly, this chaos coexists with the LC1:2 which is stable up to the 1:2TR-line (dashed green line in Fig. 3 map) where it converts to inhomogeneous torus. The inhomogeneous torus is stable over a narrow interval of coupling strength while continuing the coexistence with the LC5:5 generated chaos. The return times maps (Fig. 7) illustrate the process of the inhomogeneous torus destruction which occurs inside the interval of Q from 0.461 to 0.468 in Fig.6 where the LE value gradually increases from zero indicating growing chaos which coexists with, then jumps to merge with the chaos generated by the evolution of LC5:5.

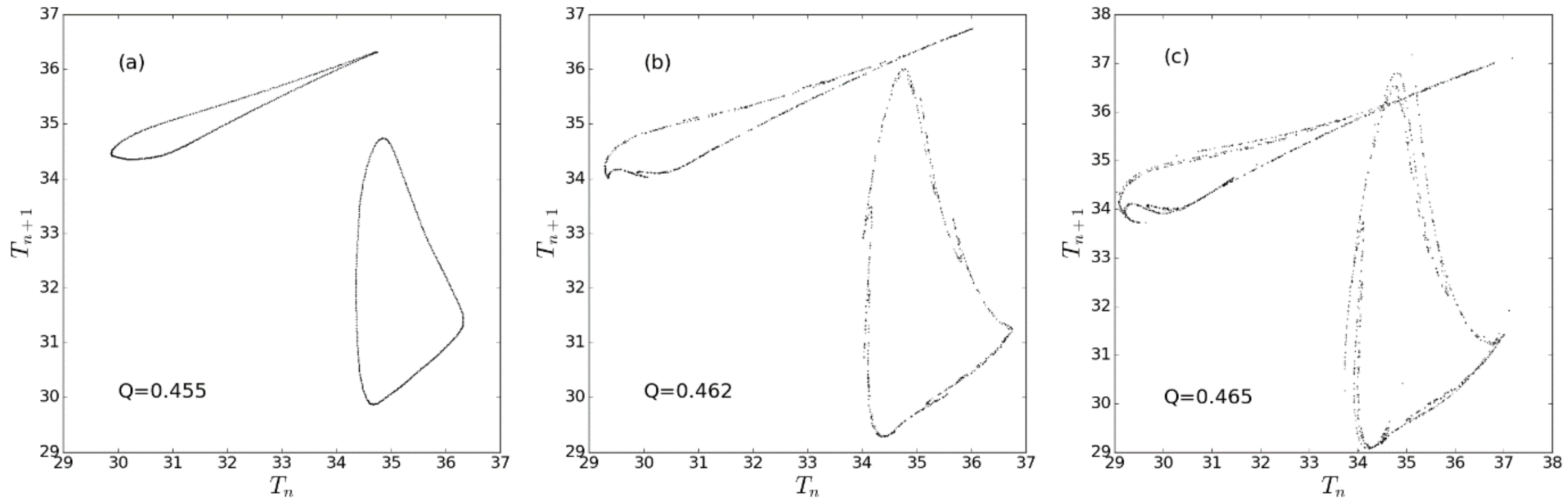

**Fig. 7** Return times map T(n+1) vs T(n) calculated for **α**=5000 using Poincare section of trajectory of the small amplitude oscillator demonstrating the inhomogeneous torus destruction which begins at Q = 0.461. (a) shows torus, (b) and (c) show torus destruction producing growing chaos.

The LE graph in Fig 8 shows coexistence of LC1:2 and chaos at both ends of the LC1:2's long stable span, including doubling to LC2:4, for α = 2000. The chaos originates from the period doubling cascades of the LC5:5, and presumably would be continuous across the region if not for the existence of the LC1:2. We assume that the extent of the chaos from both directions indicates the point at which the basin of attraction of the LC1:2 captures all phase space.

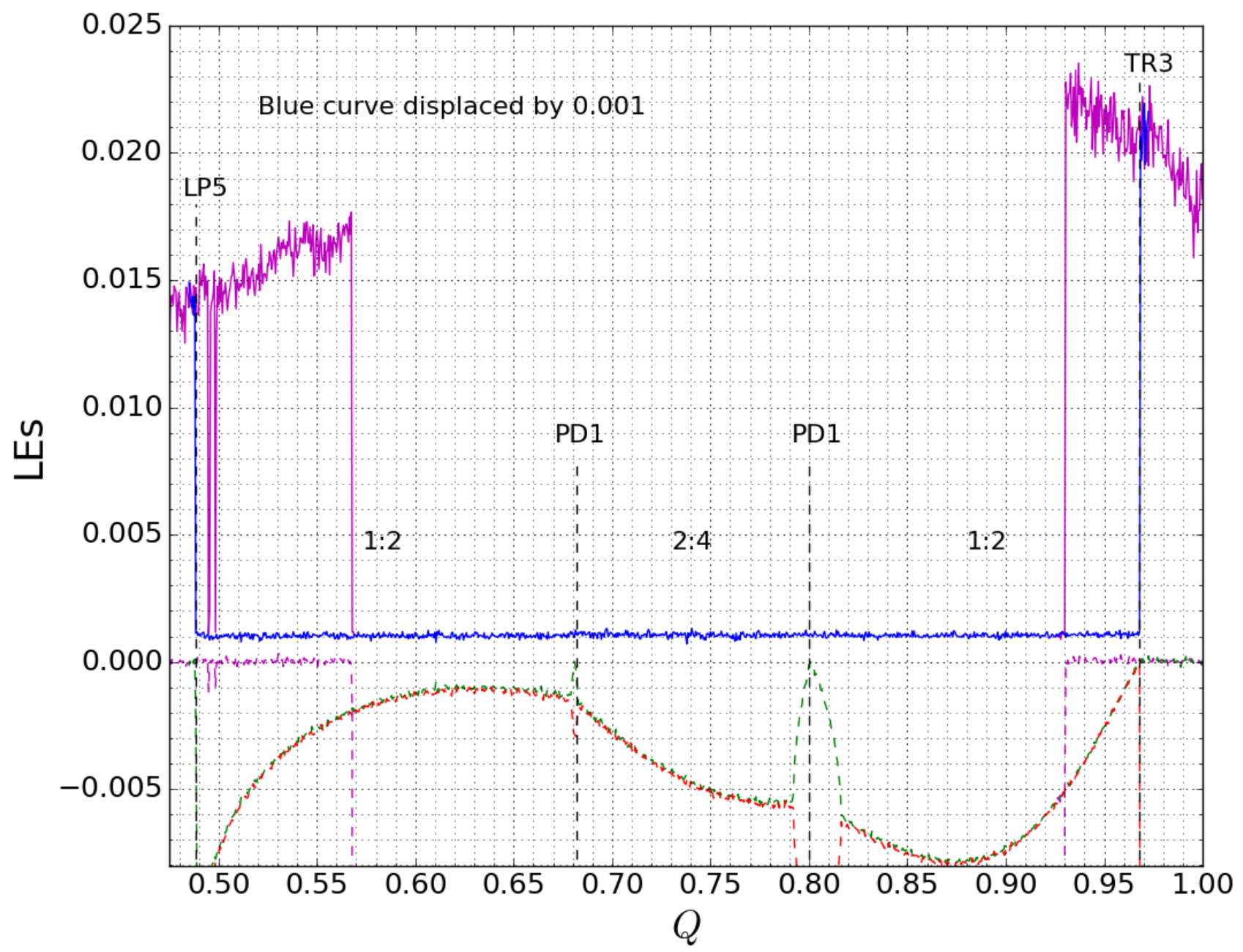

Fig 8. *LEs for α = 2000 starting from LC1:2 (blue) and from chaos (maroon), demonstrating the extensive regions of coexistence*.

An interesting appearance of additional highly asymmetric limit cycles is associated with the 1:2 torus bifurcation surrounding much of the left chaotic region indicated in Fig. 3. The formation of inhomogeneous resonant LCs from this torus may be suspected, and they were identified by direct numerical integration. The further bifurcation analysis by AUTO continuations results in the LC*n*:2*n* seen in Fig. 3 where the Arnold's tongues with IHLCs 5:10, 7:14, 8:16, 9:18 are displayed. Although the lines of the saddle-node bifurcations cover a large area in the parameter space, these IHLCs are stable only near the tips of the tongues, because they are period doubled to chaos if either of the parameters are shifted from the tips. The discovery of other IHLCs originated on 1:2TR is difficult because a long portion of the 1:2TR-line is masked by LC5:5. Another problem is that extremely narrow Q-intervals have to be checked and very long trajectories have to be analyzed to find overlapping IHLCs with high period. For example, for α=7000 direct calculations show many IHLCs (11:22, 12:24, 15:30, 16:32, etc) and we assume that other IHLCs with rotation numbers n:2n are invisible only because of the limited resolution of calculations. The formation of this family of IHLCs is classical dynamics on an inhomogeneous torus leading to chaos.

## Conclusion and Discussion.

The emergence of multistability and collective dynamical regimes with broken symmetry in populations of identical oscillators have been of interest for several decades, especially in the study of biological phenomena [[40, 41] and references therein]. Although the choice of model for the single oscillator has influence on the formation of nontrivial oscillating regimes in homogeneous populations, the type of coupling is crucial to the formulation of the regimes, even for oscillators with simple two-dimensional limit cycles. For example, [42] demonstrated the arising of oscillations with broken symmetry in a system of two directly coupled Stuart-Landau oscillators with combined attractive and repulsive couplings. [43] observed the symmetry-breaking transition from homogeneous to inhomogeneous limit cycle in coupled identical Stuart-Landau or van der Pol oscillators under mean-field coupling with an additional filter in the self-feedback path.

Here we considered the manifestations of multistability and asymmetry of collective regimes in the model of two identical 3-dimensional Repressilators indirectly coupled via production and diffusion of signal molecules, paying special attention to the joint influences of coupling intensity and internal parameters of the isolated oscillators. The Repressilator is a standard ring oscillator composed of three nonlinear elements (genes) which produce three repressors resulting in a stable limit cycle. Here we assume that the rates of transcription and the degree of repression (Hill coefficients) for all genes are identical while the degradation of one repressor (A in system Eq(1)), which is not correlated with the production of autoinducer and is not activated by it, is several times faster than the degradation of the other two repressors (B, C). Our previous works [34, 35] found the intervals of other model parameters which provide the rich multistable dynamics in system Eq.(1) allowing us to focus here on the origin of broken symmetry.

As a starting point for analysis we carried out the bifurcation continuations of steady states and IP and AP limit cycles originated from corresponding Andronov-Hopf bifurcations as a function of coupling strength under fixed values of $\boldsymbol{\alpha}$ and β's = 0.5,0.1,0.1 (see Section 3). Similar to previous results [34, 35] the bifurcation analysis of APLC revealed two Neimark-Sacker bifurcations which control the stability of APLC and an $\boldsymbol{\alpha}$-dependent pitchfork bifurcation in the low-Q region which produces IHAP characterized by nearly identical amplitudes with phase shift moved away from 180°. Q-continuation of the unstable IPLC demonstrates the appearance of an unstable asymmetric branch in the high-Q region which has been continued to its stabilization as inhomogeneous limit cycle with very different amplitudes LC1:2 (Fig.2). The region of its stability is limited by torus and saddle-node bifurcations which may occur sequentially as coupling strength changes. To our knowledge this route from IPLC to inhomogeneous limit cycles with vastly different amplitudes has never been observed for coupled identical oscillators. Additional highly asymmetric limit cycles are created by the 1:2 torus producing a family of inhomogeneous resonant LCs with rotation n:2n.

Q-continuation of the unstable HSS finds pitchfork bifurcation to IHSS which are continued to their stabilization by saddle-node bifurcations. This attractor, frequently named in literature as “oscillation death”, coexists with homogeneous steady state and can produce inhomogeneous limit cycle under appropriate values of other parameters as shown in previous studies of quorum-sensing coupled Repressilators [23-25] and confirmed in electronic version of the model [34]. This mechanism for emergence of IHLC has been known for a long time [44] and its new realizations for more complex coupling schemes have been recently published [43]. A distinguishing property of these IHLCs is their rotation around different steady states, in contrast to the 1:2LC described in this work.

Bifurcations of these discovered attractors for β’s = 0.5,0.1,0.1 were used to calculate the two-parameter (Q-α)-phase diagram Fig. 3 over a very extended area of **α** which contains the main collective attractors. Keeping in mind that our focus here is on the emergence of asymmetrical regimes, the map omits unrelated regimes. The torus bifurcation of APLC and IHAPLC makes a closed line which is the border of the region of complex dynamics. Many resonant cycles are formed on the torus with classical Arnold’s tongues near the torus boundary. We present here only the limit cycle with rotation number 5:5 for which parameter continuations found three different versions, one of which creates chaos via a period doubling cascade over a vast region of this map.

Contribution from IPLC bifurcations to this island of complex dynamics is indicated by the large amount of inhomogeneous cycle 1:2 in the map. The LC1:2 is another source of chaos, resulting from two mechanisms: first, the destruction of inhomogeneous torus and, second, a period doubling cascade of LC1:2 in the central part of the phase diagram. A typical example of joint chaotization of LC5:5 and LC1:2 when coupling strength increases has been presented in Section 3 (Fig. 6). Although LC1:2 is unstable after torus bifurcation it remains important in the formation of inhomogeneity in the chaos skeleton because the values of its multipliers are close to unity over large intervals of parameters thereby allowing long inhomogeneous pieces in the chaotic time series. The chaos in central island is composed of asymmetric pieces, which due to switching results in symmetric chaos.

The asymmetric limit cycle generated by pitchfork bifurcations of the IP limit cycle is structurally different from other regimes, and this difference allows their coexistence in 8-dimensional phase space. An example are the large regions with coexistence of 1:2LC and symmetric chaos indicated in Fig. 8 where the symmetric chaos formed from 5:5 doubling extends significantly into the LC1:2 regions. Bistability of 5:5LC and 1:2LC exists in several parameter regions and both attractors can cooperate to produce chaos via period doubling or inhomogeneous torus destruction (see e.g. Fig. 6). Although we concentrate on the presentation of different regular attractors and their bifurcations, it is appropriate to restate that most of the volume in (Q-α)-parameter space has chaos as its sole oscillatory dynamic as suggested by the LE plot in Fig. 4. This prevalence of chaos is unusual for coupled identical simple oscillators such as the ring ones used here. It is interesting that although the APLC is prevalent for the coupling used in the model system (1), while stable IPLC makes no appearance at all for βs=(0.5,0.1,0.1), both these limit cycles make substantial contributions to the system’s complex dynamics via their bifurcations.

To demonstrate the unusual abilities of the indirect quorum-sensing mean field coupling formalized in system (1) to generate well developed multistability and symmetry breaking in two coupled identical Repressilators over very wide intervals of model parameters, we presented the (Q-α) phase diagram of collective regimes which seem to us the most interesting over significant areas of parameter space. However, in this model there are many other attractors, e.g. the large family of long resonant cycles, the set of periodic windows, and both these types of cycles may exist in symmetric and asymmetric forms which are outside the scope of the current work.

We believe that the results obtained here can be important to the understanding of multistable collective symmetric and asymmetric modes in other multidimensional oscillators coupled indirectly by a mean-field coupling like the quorum-sensing mechanism in system Eqs (1).

**Acknowledgments**
This work is partially supported by a grant from the Russian Foundation for Basic Research 19-02-00610 as well as the Government assignment to the Lebedev Physical Institute (Project No. 0023-2019-0011).
We thank N. Stankevich and A. Kazakov for their interest to our work.

The data that supports the findings of this study are available within the article.